\documentclass[aps,pra,twocolumn,showpacs]{revtex4}
\usepackage{graphicx}
\begin{document}
%\sloppy
%\draft
\title{ Single point potentials with total resonant tunneling 
}
\author{A.V. Zolotaryuk}
\affiliation{Bogolyubov Institute for Theoretical Physics, National Academy of
Sciences of Ukraine, Kyiv 03680, Ukraine}

\date{\today}

\begin{abstract}
Two rectangular models described by the one-dimensional Schr\"{o}dinger equation with sharply localized potentials are suggested. The potentials have a multi-layer thin structure being composed from adjacent barriers and wells. Their peculiar tunneling properties are studied in considerable detail. Particularly, in the zero-range limit when the potentials are squeezed to a single point, sharp peaks with total transmission are observed 
at certain (positive and negative) quantized values of the potential strength constant forming infinite discrete sets.  Beyond these sets, the barrier-well structures behave as a perfectly reflecting wall. The transcendental equations with respect to potential strengths, the solutions of which determine transmission (resonance) sets, are derived. In this regard, both the models are exactly solvable. 
  The energy dependence of an incident particle is shown to reveal a resonance behavior, being completely different from that observed in a typical double barrier structure.
\end{abstract}

\maketitle

\section{Introduction}

The mechanism of resonant or non-resonant tunneling through two or more successive potential barriers plays an important role in studying different phenomena in quantum physics and optics including various aspects of engineering at micron and nanoscale levels \cite{te}. Oriented towards the investigation of the resonant tunnel effect, beyond the standard case of double barrier (DB) penetration, it is of interest to search for the models where other mechanisms of resonant transmission can take place. Thus, a one-dimensional model of resonances described by 
the Schr\"{o}dinger equation with Dirac's delta function barrier and an abrupt mass jump has been suggested and studied in Ref.~\cite{a-n}. It is worth mentioning that the motion of a particle with a position-dependent mass under different conditions (for instance, including an oscillating field) with further applications to nanostructures was a main object of studies in recent publications \cite{ll}.
Another example of recent studies relevant to resonant tunneling are one-dimensional
sharply localized potentials consisting of barriers with adjacent wells \cite{c-g,tn,zci2}. Here the approach is based on the search for those finite-range potentials which in the zero-range limit exhibit peculiar features like a non-zero transmission under certain conditions. 

In general, the models with zero-range potentials that describe point or contact interactions (for details and references see books \cite{a-h,alku}), are widely used in various applications to atomic physics \cite{do}, condensed matter systems 
\cite{ael}, acoustic and electromagnetic waveguide networks \cite{fjk}. Currently, there is a growing interest in the quantum dynamics of particles in structures of low dimension which can be described by graphs (for details see survey \cite{ku}) using different couplings and their approximations at the graph vertices 
\cite{exner,ftc,m}. Of particular importance are nanoscale quantum devices such as thin quantum waveguides \cite{acf} and potential-controllable spectral filters 
\cite{tc}. 

 The point interaction models admit exact closed analytical solutions, which are rare but quite useful. They provide relative simple situations, where an appropriate regularization procedure can be chosen to be in relevance with real structures being important for technological applications. Thus, remarkable structural analogies between the Dirac delta function and Coulomb potentials in one dimension have been discussed 
\cite{b}. In the case of moving this delta potential with a constant velocity, the exact propagator has been evaluated 
 by summing over the spectrum of eigenstates and by calculating the path integral directly \cite{cl}. In a number of papers \cite{ckn}, the relativistic case of a delta function potential in the one-dimensional Dirac equation has been treated.
On the other hand, for a general class of point interactions, Green's functions have been computed by using the reflection and transmission coefficients 
\cite{scl}. The Schr\"{o}dinger equation with   
 two- and three-dimensional delta function potentials has been the object of considerable study \cite{h}. In particular, exactly solvable models has been elaborated in two and three dimensions by renormalizing the kinetic energy operator \cite{pc}. It is also worth to mentioning the recent studies of time aspects of point interactions \cite{km}.

In this paper, we would like to address the problem of resonant tunneling through single point potentials. The regularized (finite-range and realistic) structure of these potentials has the form of barriers with adjacent one or more wells. The presence, at least, one well will be shown below to play a crucial role in the mechanism of tunneling. The key point of our approach is the understanding of this mechanism which appears to differ from that well known in DB structures. A small region of large repulsive potential (barrier), which is followed by a small region of large attractive potential (well), can be served as an interesting example in one dimension. Under an appropriate parametrization of parameters, this barrier-well (BW) potential in the zero-range limit becomes the derivative of Dirac's delta function referred to as a point dipole studied in numerous publications 
\cite{s,cnp}. 
Another example, which can be treated analytically, is the second derivative of a delta function but with a renormalized strength constant \cite{zci2}, which essentially reduces the singularity at the origin of the point interaction. Below, it will also be demonstrated that the zero-range idealization is an useful prompting for investigation of tunneling through finite-range structures.

\begin{figure}
%\begin{center}
%\begin{center}
\centerline{\includegraphics[width=0.5\textwidth]{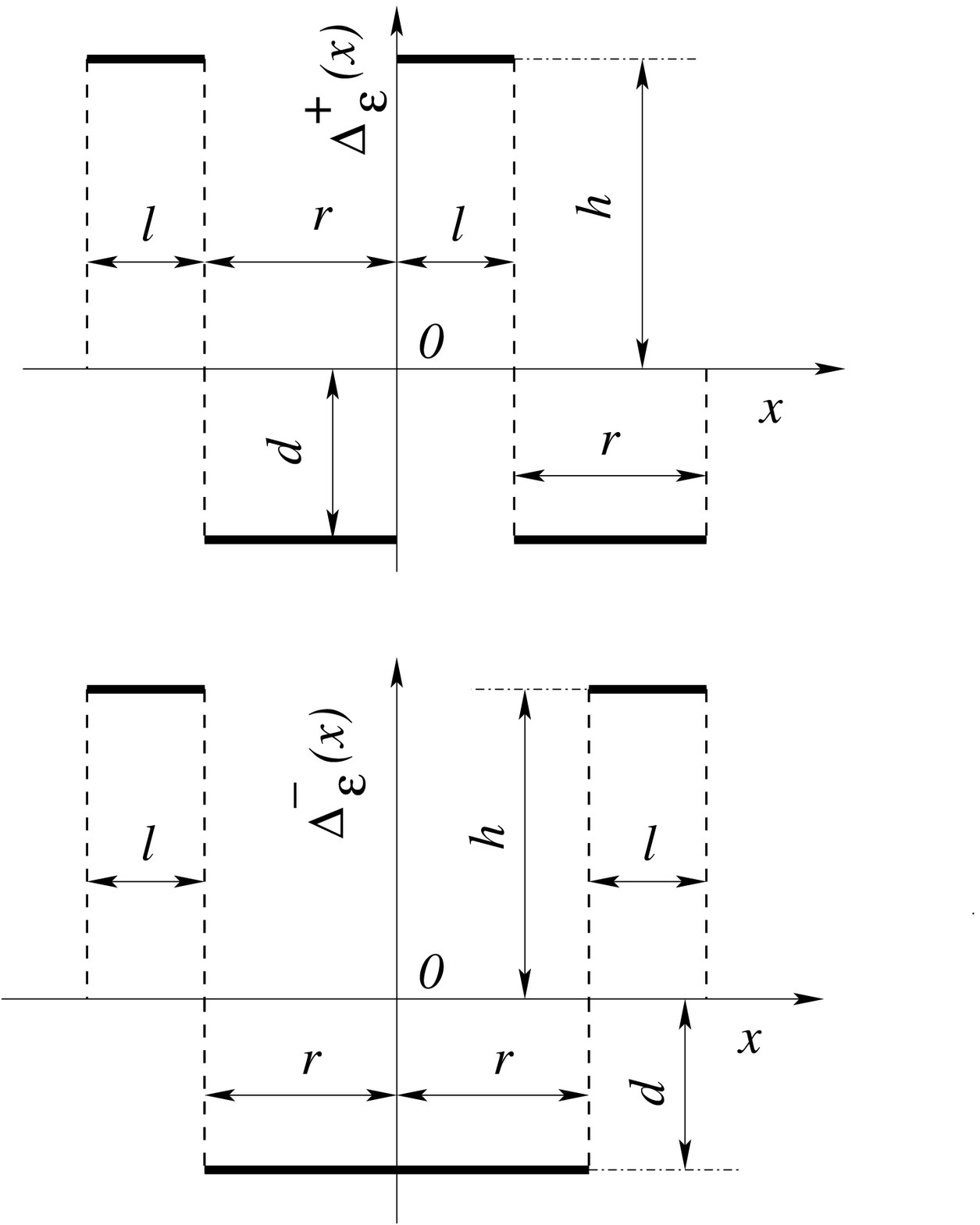}}
%\end{center}
\vspace{2pt}
\caption{ Rectangular models: $\Delta_\varepsilon^+(x)$ (top) and 
$\Delta_\varepsilon^-(x)$ (bottom) defined according to Eqs.~(\ref{2}) and 
(\ref{2a}), and parametrized by Eqs.~(\ref{3})  with $\sigma^+ =\sigma^- =1$.}
%\end{center}
\label{fig.1}
\end{figure}

Thus, we consider the systems that present two successive barriers with a well inserted between them or a barrier with adjacent wells located at both its sides. 
More precisely, these systems are supposed to be described by the following two potentials (see Fig.~\ref{fig.1}): 
\begin{eqnarray}
\Delta^+_\varepsilon (x) &=& \Delta'_\varepsilon (x+r) + \Delta'_\varepsilon (x-l) 
\nonumber \\ 
\Delta^-_\varepsilon (x)& =& \Delta'_\varepsilon (x+r) + \Delta'_\varepsilon (r-x),
\label{2}
\end{eqnarray}
 given through the rectangular function $\Delta'_\varepsilon (x)$ of the BW form defined by the equations
\begin{equation}
 \Delta'_\varepsilon (x) = \left\{ \begin{array}{ll} 
~~ h & \mbox{for}~~ -l < x< 0, \\
-d  & \mbox{for} ~~ 0 < x< r , \\
~~  0 & \mbox{otherwise} .
\end{array} \right.
\label{2a}
\end{equation}
Here, $h$ and $l$ are height and width of the barrier, and $d$ and $r$ are depth and width of the well, respectively.
 These parameters can be expressed through a squeezing parameter $\varepsilon$ as follows
\begin{eqnarray}
h&=& { 2 \sigma^+ \over c_1(c_1 + c_2) } \, 
\varepsilon^{-2},~~ ~l= c_1 \varepsilon, \nonumber \\
d &=& { 2\sigma^- \over c_2 (c_1 +c_2)}  \, \varepsilon^{-2}, 
~~~r= c_2  \varepsilon,
\label{3}
\end{eqnarray}
with arbitrary positive constants $c_1$ and $c_2$. The constants $\sigma^\pm$ are defined as follows: $(\sigma^+, \, \sigma^-) =(1, \, 1)$ for $\alpha =0$, 
$(\sigma^+, \, \sigma^-)= (1, \, \sigma) $ for $\alpha >0$, and  
$(\sigma^+, \, \sigma^-) =(\sigma, \, 1)$ for $\alpha < 0$, where $\sigma$ is an arbitrary positive parameter. This definition means that the parameter $\sigma$ is assigned only to the wells. Parametrization (\ref{3}) has been chosen in such a way that in the particular case $\sigma =1$, one obtains the weak limit:  
 $\Delta'_\varepsilon (x) \to \delta'(x) \doteq d\delta(x) / dx$ in the sense of distribution theory as $\varepsilon \to 0$. In this case,  $lh =rd$, i.e., the condition $\int_{-\infty}^\infty \Delta'_\varepsilon (x)dx =0$ is fulfilled. 
The parameters $\sigma^\pm$ (or $\sigma$) have been incorporated to investigate a more general class of point interactions than the $\delta'$-potential. They control the deviation of the standard $\delta'(x)$ distribution from the zero-range limit of the whole $\sigma$-family of point interactions which cannot be defined as distributions. Furthermore, the parameter $\sigma$ appears to be useful in studying 
the situation, when the wells continuously disappear resulting in a DB structure as $\sigma \to 0$. In the particular case $\sigma = 1$, within parametrization 
(\ref{3}), one can prove that  
  $\Delta^\pm_\varepsilon(x) \to \delta'(x+0) \pm  \delta'(x-0)$ in the sense of distributions defined on the space of appropriately chosen test functions.
Therefore, if $\sigma = 1$, the potentials $\Delta^\pm_\varepsilon(x)$ may be called  double $\delta'$-like structures or one-dimensional double point dipoles (or double plane capacitors).

Thus, we consider the one-dimensional Schr\"{o}dinger equation for stationary states, which in the limit that neglects the interaction between electrons, reads  
\begin{equation}
-\psi''(x) + \alpha  \Delta_\varepsilon^\pm(x) \psi(x) =E \, \psi(x),
\label{1}
\end{equation}
where $\psi(x)$ is the wave function
for a  particle of mass $m$ and energy $ E $ (we use units in which $\hbar^2/2m =1$), and $\alpha$ is an interaction strength constant. 

 Recently, 
using various regularization schemes \cite{c-g,tn,z} for the distribution $\delta'(x)$, the potential $\alpha \delta'(x)$ inserted in Eq.~(\ref{1}) instead of $\alpha \Delta_\varepsilon^\pm(x) $, has been shown to be partially transparent at a certain set of quantized values for $\alpha$, but acting as a perfect reflecting wall for the values different from this transmission set. This resonant transmission property has rigorously been proved in Ref.~\cite{gm} by using a general class of regularizing sequences. However, the partial transparency decreases very rapidly with the growth of $\alpha$. Consequently, from a practical point of view, it would be of interest to search the single point interaction systems that admit a total resonant transmission. A composition, at least, of two adjacent $\delta'$-function potentials is expected to permit a total resonant behavior. That is why the two double $\delta'$-like structures described by Eqs.~(\ref{2}) and plotted in Fig.~\ref{fig.1} have been chosen for this purpose. In the zero-range limit, for one of these structures the two point dipoles are arranged forward (top of Fig.~\ref{fig.1}), whereas for the other one they are arranged oppositely (bottom of Fig.~\ref{fig.1}).

\section{Finite-range solutions for the $\Delta_\varepsilon^\pm(x)$ models}

The positive energy solution of the Schr\"{o}dinger equation (\ref{1}) with potentials (\ref{2}) can be written through the transfer matrices $\Lambda^\pm $, connecting the boundary conditions for the wave function 
$\psi(x)$ and its derivative $\psi'(x)$ at $x=x_1=-l -r $ and $x=x_2 =l+ r $: 
\begin{eqnarray}
\left( \begin{array}{cc} \psi(x_2)  \\
\psi'(x_2) \end{array} \right) 
 = \Lambda^\pm \left(
\begin{array}{cc} \psi(x_1)   \\
\psi'(x_1)   \end{array} \right), ~~ \Lambda^\pm =  \left(
\begin{array}{cc} \lambda_{11}^\pm~~ 
\lambda_{12}^\pm \\
\lambda_{21}^\pm ~~ \lambda_{22}^\pm \end{array} \right) .
\label{4}
\end{eqnarray}
This approach appears to be very useful because the reflection and transmission coefficients can be computed immediately in terms of the transfer matrix. On the other hand, the zero-range limit of transfer matrices directly leads to the corresponding connection matrix, on the basis of which all possible (four-parameter) point interactions in one dimension can be classified \cite{adk}. Similarly, the scattering data can easily be calculated through the connection matrix (see, for instance, Refs.~\cite{tn,scl,cnp}).
  
Thus, by matching the values and derivatives of the wave function $\psi(x)$ at each potential discontinuity, after straightforward calculations we arrive at 
\begin{eqnarray}
\lambda^+_{11} &=& 
\cos(2pl) \cos^2(qr) \nonumber \\
&-& \frac14  \left(3\, {p \over q} +{q \over p}\right) \sin(2pl) \sin(2qr)  \nonumber \\
&+& \left[ {p^2 \over q^2}\, \sin^2(pl) - \cos^2(pl) \right] \sin^2(qr), 
\nonumber \\
\lambda^+_{12} &=& 
\frac1p \sin(2pl) \cos^2(qr) +\frac1q \cos^2(pl)\sin(2qr) \nonumber \\
&-& \left( {p \over q} +{q \over p} \right) \left[ \frac1p \sin(pl) \cos(qr)
\right. \nonumber \\
&+& \left. \frac1q \cos(pl) \sin(qr)\right] \sin(pl) \sin(qr), \nonumber \\
\lambda^-_{21} &=& 
-p \sin(2pl) \cos^2(qr)-q \cos^2(pl) \sin(2qr) \nonumber \\
&+& \left( {p \over q} + {q \over p} \right) 
\left[ p \sin(pl)\cos(qr) \right. \nonumber \\
 &+& \left. q \cos(pl)
\sin(qr) \right]\sin(pl) \sin(qr), \nonumber  \\
\lambda^+_{22} &=& 
\cos(2pl) \cos^2(qr) \nonumber \\ 
&-& \frac14  \left({p \over q} +3 \, {q \over p} \right) \sin(2pl) \sin(2qr)  \nonumber \\
&+& \left[ {q^2 \over p^2} \sin^2(pl) - \cos^2(pl) \right] \sin^2(qr)
\label{5}
\end{eqnarray} 
and
\begin{eqnarray}
\lambda^-_{11} &=& \lambda^-_{22} =
\cos(2pl) \cos(2qr)\nonumber \\ 
& -&\frac12  \left({p \over q} +{q \over p}\right) \sin(2pl) \sin(2qr), 
 \nonumber \\
\lambda^-_{12} &=& 
\frac1p \sin(2pl) \cos(2qr) \nonumber \\
&+&\frac1p \left[ {p \over q}\cos^2(pl) - {q \over p}\sin^2(pl)\right] 
\sin(2qr), \nonumber \\
\lambda^-_{21} &=& 
-p \sin(2pl) \cos(2qr) \nonumber \\
&+& p \left[ {p \over q}\sin^2(pl) - {q \over p}\cos^2(pl)\right] \sin(2qr),
\label{6}
\end{eqnarray} 
with
\begin{equation}
 p \doteq \sqrt{E-\alpha h }\, ,~~ q \doteq \sqrt{E +\alpha d }  \, ,
 ~~E \doteq k^2.
\label{7}
\end{equation}
Here, $p$ and $q$ can be either real or imaginary depending on the sign and values of $\alpha$ and the wave number $k$.

According to parametrization (\ref{3}), the quantities $p$ and $q$ are divergent as $\varepsilon \to 0$, while the products 
$pl$ and $qr$ must be constant in this limit. Under the latter assumption, it follows from the form of Eqs.~(\ref{5}) and (\ref{6}) that the matrix elements $\lambda_{11}^\pm$ and $\lambda_{22}^\pm$ are constant, $\lambda_{12}^\pm \to 0$, and $\lambda_{21}^\pm$ is divergent as $\varepsilon \to 0$. However, one can expect that under some appropriate conditions, the cancellation of divergences may happen. Indeed, the terms $\lambda_{21}^\pm$ can be factorized as
\begin{eqnarray}
\lambda^+_{21} &=& \left[ \left( {p \over q} +{q \over p} \right) \sin(pl) \sin(qr)
-2\cos(pl) \cos(qr) \right] \nonumber \\ &\times&
\left[p \sin(pl) \cos(qr) + q \cos(pl)\sin(qr) \right] ,
\label{a} \\
\lambda^-_{21} &=& 2 \left[ {p \over q} \sin(pl)\sin(qr) - \cos(pl) \cos(qr) \right]
\nonumber \\ 
&\times& \left[ p \sin(pl)\cos(qr) + q \cos(pl) \sin(qr)  \right] .
\label{b} 
\end{eqnarray}
Using next this representation and imposing the constraint
 $\lambda^\pm_{21}=0$, we find the conditions under which the divergences cancel in the $\varepsilon \to 0$ limit. As a result, for the potential $\alpha \Delta^+_\varepsilon(x)$ one can write the following two equations:
\begin{eqnarray}
 &&  2\cos(pl) \cos(qr) = \left( {p \over q} +{q \over p} \right)
\sin(pl) \sin(qr) ,  \label{8} \\ 
&& p \, \sin(pl) \cos(qr) =-  q \, \cos(pl)\sin(qr).  \label{9}
\end{eqnarray}
Similarly, for the potential $\alpha \Delta^-_\varepsilon(x)$, the constraint $\lambda^-_{21}=0$ also leads to the two equations: the same Eq.~(\ref{9}) and
\begin{equation}
 p \, \sin(pl) \sin(qr) = q \, \cos(pl) \cos(qr)  .
\label{10} \\
 \end{equation}

Next, imposing the condition $\lambda_{21}^\pm =0$ and using Eqs.~(\ref{8})-(\ref{10}), one can transform the rest of the matrix elements
in Eqs.~(\ref{5}) and (\ref{6}), yielding the representations
\begin{eqnarray}
\Lambda^+_1  &=&  -I,  \label{11} \\
\Lambda^+_2 &=&  \left(
\begin{array}{cc} {\cos^2(pl) \over \cos^2(qr) }~~~~~ 
2 \left[ {\cos^2(qr) \over p \sin(2pl) } + 
{\cos^2(pl) \over q \sin(2qr) }  \right]  \\
0 ~~~~~~~~~~~~~~~~~~~ {\cos^2(qr) \over \cos^2(pl) }~~~~~~ \end{array} \right)\! ,
\label{12} \\
\Lambda^-_1 &=&  -\Lambda^-_2 \, = \, \left(
\begin{array}{cc}\!\!\! -1~~~~  { 2 p^{-1} (p^2 -q^2) \tan(pl) \over 
p^2 \tan^2(pl) +q^2 }  \\
~~0 ~~~~~~~~~~~~~~~ -1~ ~~~~~~~\end{array} \right) .
\label{13} 
\end{eqnarray}
where $I$ is the unit matrix. Here the representations $\Lambda_1^+$ and $\Lambda_2^+$ for the $\alpha \Delta_\varepsilon^+(x)$ potential have been obtained under conditions (\ref{8}) and (\ref{9}), respectively. Similarly, the representations $\Lambda_1^-$ and $\Lambda_2^-$ for the $\alpha \Delta_\varepsilon^-(x)$ potential correspond to Eqs.~(\ref{10}) and (\ref{9}), respectively. 

\section{Transmission sets in the zero-range limit}

In this section, we analyze the $\varepsilon \to 0$ limit of 
Eqs.~(\ref{8})-(\ref{10}) as well as $\Lambda^\pm_{1,2}$-matrices (\ref{12}) and (\ref{13}). As a result, using the asymptotics
\begin{equation}
p  \to {\rm i}  \, \sqrt{ 2\alpha \sigma^+ \over c_1(c_1 +c_2) }~  \, \varepsilon^{-1}~~  
\mbox{and}~~  q  \to  \sqrt{2\alpha \sigma^- \over c_2(c_1 +c_2)  } 
\,~ \varepsilon^{-1} \, , 
 \label{15}
\end{equation}
 for the potential $\alpha \Delta^+_\varepsilon(x)$ with any $\alpha \in (-\infty, \, \infty)$, one finds that the $\varepsilon \to 0$ limit of Eq.~(\ref{8}) yields
the equation
\begin{equation}
 \left( \sqrt{b \sigma^- \over \sigma^+} - \sqrt{\sigma^+ \over b \sigma^-}~ \right)  \tanh \!\sqrt{ 2\alpha \sigma^+ \over 1 + b^{-1} } \, \, 
\tan \! \sqrt{ 2\alpha \sigma^- \over 1 + b} =  2   ,
\label{17}
\end{equation}
 where $b = c_1 / c_2 $. Similarly, for the potential $\alpha \Delta^-_\varepsilon(x)$,  we get the $\varepsilon \to 0$ limit of Eq.~(\ref{10}):
\begin{equation}
\tanh \! \sqrt{ 2\alpha \sigma^+ \over 1 + b^{-1} } \, \, 
\tan \! \sqrt{ 2\alpha \sigma^- \over 1 + b} =
 - \sqrt{b \sigma^- \over \sigma^+} \, \,  .
\label{19}
\end{equation}
 Equations~(\ref{17}) and (\ref{19}) admit an infinite but countable set of solutions (roots), which we denote by 
$ \alpha_n^\pm ~ (n = 0, \, \pm 1, \, \ldots )$ for 
the potentials $\alpha \Delta_\varepsilon^\pm (x)$, respectively. 
In its turn, Eq.~(\ref{9}) also determines a countable set of quantized values for $\alpha$, being the same for both the potentials 
$\alpha \Delta_\varepsilon^\pm (x)$. The corresponding limiting equation becomes 
\begin{equation}
\tanh\! \sqrt{ 2\alpha \sigma^+ \over 1 + b^{-1} } = \sqrt{b \sigma^- \over \sigma^+} \, \tan \!\sqrt{ 2\alpha \sigma^- \over 1 + b}  ~.
 \label{21}
\end{equation}
We denote this set, which in the particular case $\sigma^+= \sigma^- =1$ coincides with the corresponding resonance set found in Ref.~\cite{z} for the $ \delta'$-potential,
 by $ \alpha'_n  ~(n = 0, \, \pm 1, \, \ldots)$. However, the zero-range limits of the $\Lambda_2^\pm$-matrices described by Eqs.~(\ref{12}) and (\ref{13}) are different. Since the off-diagonal matrix elements $\lambda_{12}^\pm$ in these equations disappear as $\varepsilon \to 0$,  the connection matrices 
$\Lambda_{1,2}^\pm$ take the following final form:  
\begin{equation}
\lim_{\varepsilon \to 0}\Lambda_1^\pm  = -I ,~~
\lim_{\varepsilon \to 0} \Lambda^+_2  = {\Lambda'}^2,~~
 \lim_{\varepsilon \to 0}\Lambda_2^-  = I ,
\label{24}
\end{equation}
where the $\Lambda'$-matrix is given by 
\begin{equation}
 \Lambda' =  \left(\!
\begin{array}{cc}  \theta_n ~~~0 ~~ \\
\! \! \!0 ~~~\theta_n^{-1}\!\! \! \end{array} \!\! \right), ~~\theta_n \doteq 
{ \cosh \! \sqrt{2\alpha'_{n} \! \sigma^+ /(1+b^{-1})} \over
\cos\!\sqrt{2\alpha'_{n} \sigma^- /(1+b)} } \neq 1.
\label{24a}
\end{equation}
In the particular case $\sigma^+ =\sigma^- =1$, this connection matrix corresponds to the $\delta'$-potential computed in Refs.~\cite{c-g,z,gm}.

Note that the existence of Eqs.~(\ref{17})-(\ref{21}), which determine the conditions for a non-zero transmission, immediately follows from the form of the potentials $\alpha \Delta_\varepsilon^\pm(x)$. Indeed, due to the scaling given by Eqs.~(\ref{3}), in the zero-energy resonant case $E \to 0$, Eq.~(\ref{4}) becomes homogeneous and scale invariant with respect to squeezing the potentials. Therefore for the existence of non-trivial solutions, it requires a constraint on the potential parameters $\alpha, \, c_1, \, c_2$, and $\sigma^\pm$ fixing admissible  sets for $\alpha$.  In our case, the constraint can be found explicitly by matching the rescaled wave functions and its derivatives at $x/\varepsilon \doteq \xi =0, \, c_1, \, \pm c_2, \, \pm (c_1 +c_2)$. As expected, within this straightforward procedure we arrive at Eqs.~(\ref{17})-(\ref{21}).  

Thus, as follows from Eqs.~(\ref{24}) and (\ref{24a}), the point potentials determined by the zero-range limit of the 
Schr\"{o}dinger equation (\ref{1}) describe the total resonant transparency on the sets $\Sigma^\pm \doteq \{ \alpha^\pm_n \}_{n =-\infty}^\infty$ for both the potentials $\alpha \Delta_\varepsilon^\pm (x)$, respectively, and on the set 
$\Sigma' \doteq \{ \alpha'_n \}_{n =-\infty}^\infty$ for the potential 
$\alpha \Delta_\varepsilon^- (x)$, whereas only the partial transparency occurs for the potential $\alpha \Delta_\varepsilon^+(x)$ on the set $\Sigma'$.
Beyond these transmission (resonance) sets, $\lambda^\pm_{21} \to \infty$ as $\varepsilon \to 0$ and therefore Eqs.~(\ref{8})-(\ref{10}) are not valid anymore. For these continuous values of $\alpha$ (almost everywhere on the line $-\infty < \alpha < \infty$, except for the resonance sets $\Sigma^\pm$ and $\Sigma'$), both these point potentials (referred to as ``separated'' while using either the local \cite{adk} or the global \cite{ftc} description of the boundary conditions at the singularity point) are opaque. 

Note that $\Lambda'$-matrix (\ref{24a})  connects the two-sided boundary conditions at $x= \pm 0$ as
%-------------------------------24b-------------------------------------------
\begin{equation}
\psi(+0) = \theta_n \psi(-0) ~~\mbox{and}~~\psi'(-0) = \theta_n \psi'(+0),
\label{24b}
\end{equation}
%-------------------------------24b-------------------------------------------
which mean that the wave function and its derivative are discontinuous at $x=0$. As described by these equations, the discontinuity is expressed in the special way through the function $\theta_n$ defined only on the discrete set $\Sigma'$.
 In more general terms, for the one-dimensional Schr\"{o}dinger equation, the approximation of point interactions with a discontinuity of wave functions at the singularity point has been analyzed earlier in
Ref.~\cite{cs}, where instead of rectangles, the sum of three delta functions:
%-------------------------------24c-------------------------------------------
 \begin{equation}
\xi(x; a,c, \varepsilon) = c(\varepsilon) \delta(x +\varepsilon) + a(\varepsilon) \delta(x) +c(\varepsilon) \delta(x -\varepsilon) 
\label{24c}
\end{equation} 
%-------------------------------24c-------------------------------------------
has been used. Within this approach, varying the behavior of $a(\varepsilon)$ and $c(\varepsilon)$ as functions of $\varepsilon$ in a proper way, a whole class of point interactions, which realize a discontinuity of wave functions as the distance $\varepsilon $ between the delta functions tends to zero, has been analyzed.

\section{The scattering approach: Numerical considerations}

For better understanding the phenomenon of tunneling across the 
$\alpha \Delta_\varepsilon^\pm(x)$ potentials, it is reasonable to carry out
  direct numerical computations of the scattering amplitudes when the parameter $\varepsilon$ is finite. The reflection and transmission coefficients can easily be expressed in terms of the elements of the connection matrix for a given point interaction \cite{scl,cnp}. 
Similarly, for finite-range interactions, these coefficients can be computed through the transfer matrix; in our case, by employing the $\Lambda^\pm$-matrices described by Eqs.~(\ref{5})-(\ref{7}). Thus, 
omitting for a while the superscripts $``\pm"$, we consider
the scattering both from the left (labelled by the subscript $``l"$) and from the right (labelled by the subscript $``r"$) of the structures described by 
Fig.~\ref{fig.1}. As usual, we adopt the following definition for the reflection and transmission coefficients: 
\begin{equation}
\psi(x) = \left\{ \begin{array}{ll}
 {\rm e}^{{\rm i} kx} + R_l \, {\rm e}^{-{\rm i} kx}  &   \mbox{for}~~ -\infty
 < x < x_1  , \\
 T_l\, {\rm e}^{{\rm i} kx} & \mbox{for}~~~ ~x_2 < x < \infty  
\end{array} \right. 
\label{25}
\end{equation} 
from the left and 
\begin{equation}
\psi(x) = \left\{ \begin{array}{ll}
 {\rm e}^{- {\rm i} kx} + R_r \, {\rm e}^{{\rm i} kx}  &   \mbox{for}~~  
~~x_2 < x < \infty  , \\
 T_r\, {\rm e}^{-{\rm i} kx} & \mbox{for}~~~-\infty  < x < x_1  
\end{array} \right. 
\label{26}
\end{equation} 
from the right, where $x_1 = -l-r$ and $x_2 = l+r$. Inserting Eqs.~(\ref{25}) and (\ref{26}) into $\Lambda$-matrix equation (\ref{4}), we find the reflection ($R_l$ and $R_r$) and transmission  ($T_l$ and $T_r$) coefficients:
\begin{eqnarray}
R_l &=& D^{-1} \left[ \lambda_{22} -\lambda_{11} -{\rm i}\, (k \lambda_{12} +k^{-1} \lambda_{21}) \right] {\rm e}^{2 {\rm i}  k x_1}\, , \nonumber \\
R_r &=& D^{-1} \left[ \lambda_{11} -\lambda_{22} 
-{\rm i} \,(k \lambda_{12} +
k^{-1} \lambda_{21}) \right] {\rm e}^{-2 {\rm i}  k x_2}\, , \nonumber \\
T_l\, &=& T_r = 2 D^{-1} \,  {\rm e}^{ {\rm i}  k (x_1-x_2) }\, , 
\label{27}
\end{eqnarray}
where $ D = \lambda_{11} +\lambda_{22} - {\rm i} \, (k \lambda_{12} - k^{-1} \lambda_{21})$.
Using now the relation $\lambda_{11} \lambda_{22} -\lambda_{12}\lambda_{21}=1$, the probabilities of reflection (reflectivity  ${\cal R}$) and transmission  (transmissivity ${\cal T}$) can be written in the form
\begin{eqnarray}
{\cal R} &=& |R_l|^2= |R_r|^2 = {u^2 +v^2 \over 4 +u^2 +v^2 }\, , \nonumber \\
{\cal T} &=& |T_l|^2= |T_r|^2 = {4 \over 4 +u^2 +v^2 } \, ,
\label{29}
\end{eqnarray}
 where 
\begin{equation}
u = \lambda_{11} -\lambda_{22}~~~~\mbox{and}~~~~
v = k\lambda_{12} + k^{-1} \lambda_{21} \, .
\label{30}
\end{equation}
The equation for the transmissivity ${\cal T}$, where the $\Lambda^\pm$-matrices are given by Eqs.~(\ref{5})-(\ref{7}), will be used below for numerical computations by varying both the parameters $\alpha$ and $k$.

Now we focus on the direct numerical calculation of the transmissivity 
${\cal T}={\cal T}^\pm$ according to Eqs.~(\ref{29}) and (\ref{30}). The quantities $u=u^\pm$ and $v=v^\pm$ are computed by using Eqs.~(\ref{3}) and (\ref{5})-(\ref{7}). For better visualization, the logarithmic scale has been used. The case with $\sigma =1$ provides the double $\delta'$-like structure, while the case with $\sigma =0$ is relevant to the absence of wells. 

\begin{figure}
%\begin{center}
%\begin{center}
\centerline{\includegraphics[width=0.5\textwidth]{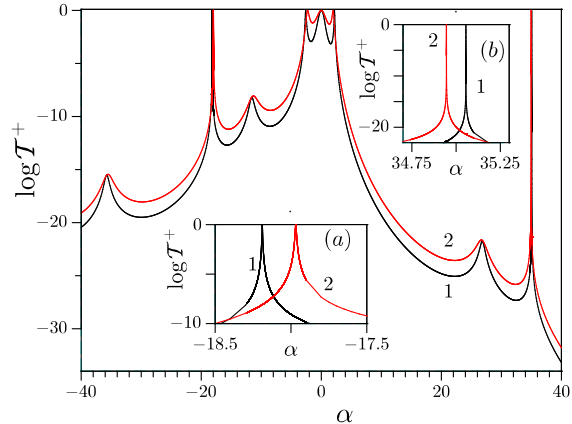}}
%\end{center}
\vspace{2pt}
\caption{ (Color online) Transmission probability ${\cal T}^+$ plotted as a function of the strength $\alpha$ at fixed wave number ($k=1$) and $\varepsilon = 0.1$ (curve 1, black) and $\varepsilon =0.2$ (curve 2, red). Five peaks with total transmission (${\cal T}^+ =1$) at $\alpha^+_{-2} =-18.26$ (inset, bottom), $\alpha^+_{-1} =-2.70$, $\alpha_0^+ =0$, $\alpha^+_{1} =2.28$, $\alpha^+_{2} =35.09$ (insert, top) and three peaks with partial transmission (${\cal T}^+ < 1$) at $\alpha'_{-2} =-35.82$, $\alpha'_{-1} =-11.66$, $\alpha'_{1} =26.87$ are clearly visible. Here $b=3$ and $\sigma =1$. }
%\end{center}
\label{fig.2}
\end{figure}

First, at fixed values of $b$ and $k$, we calculate the transmissivity ${\cal T}^\pm$ as a function of $\alpha$ for some values of $\varepsilon >0$. Thus, ${\cal T}^+$ plotted in Fig.~\ref{fig.2} as a function of $\alpha$ for two fixed values of $\varepsilon $ clearly illustrates the 
 five peaks with total transmission (${\cal T}^+ =1$) and the three peaks with partial transmission (${\cal T}^+ < 1$).
 These peaks correspond to the solutions of Eqs.~(\ref{17}) and (\ref{21}) approaching them in the $\varepsilon \to 0$ limit. In the natural (non-logarithmic) scale, these peaks appear to be very narrow even if $\varepsilon$ is not small. The location of the peaks exactly corresponds to the sets $\Sigma^+$ and $\Sigma'$. In the space between the peaks, the transmission is almost zero.  

\begin{figure}
%\begin{center}
%\begin{center}
\centerline{\includegraphics[width=0.5\textwidth]{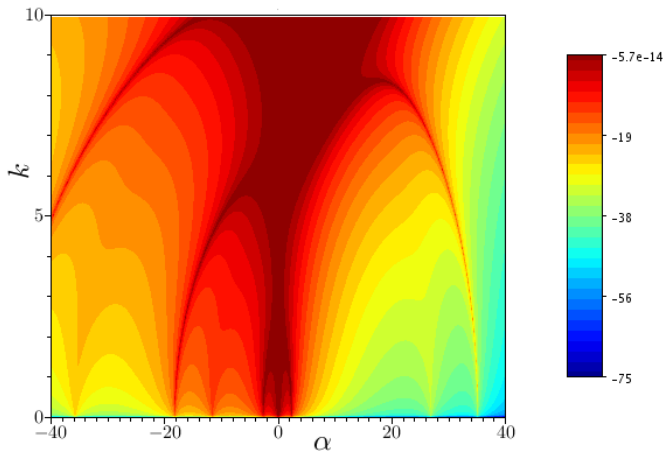}}
%\end{center}
%\vspace{2pt}
\caption{ (Color online) Transmission probability ${\cal T}^+$ 
plotted as a function of the strength $\alpha$ and the wave number $k$ with $\varepsilon = 0.2$. Here $b=3$ and $\sigma =1$. The ridges at $k \to 0$ exactly correspond to the peaks shown in Fig.~\ref{fig.2}.}
%\end{center}
\label{fig.3}
\end{figure}

\begin{figure}
%\begin{center}
%\begin{center}
\centerline{\includegraphics[width=0.5\textwidth]{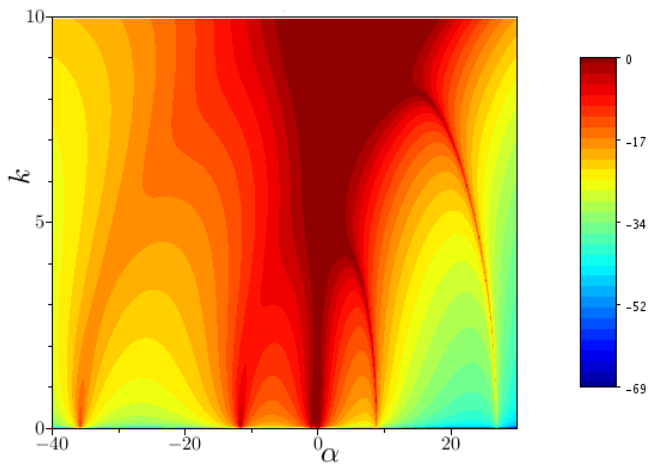}}
%\end{center}
%\vspace{2pt}
\caption{(Color online) Transmission probability ${\cal T}^-$ plotted as a function of the strength $\alpha$ and the wave number $k$ with $\varepsilon = 0.2$, $b=3$, and $\sigma =1$. The ridges at $k \to 0$ correspond to total transmission and they are given by the solutions to Eqs.~(\ref{19}) and (\ref{21}): $\alpha'_{-2}= -35.82$,  $\alpha^-_{-2}= -11.74$, $\alpha'_{-1}= -11.66$, $\alpha^-_{-1}= -1.01$, $\alpha =0$, $\alpha^-_{1}= 8.77$ and $\alpha'_{1}= 26.87$. }
%\end{center}
\label{fig.4}
\end{figure}

It is unusual that the transmission equations (\ref{17})-(\ref{21}) do not contain any dependence on the energy of an incident particle $E$ (or the wave number $k$). Therefore we have also studied the behavior of the transmission as a function of
 $k$ in a realistic case when $\varepsilon $ is not very small. To this end, we have plotted the dependence of the transmission on both the quantities $\alpha$ and $k$. In the form of contour plots, this dependence is demonstrated by Figs.~\ref{fig.3} and \ref{fig.4} for both the potentials $\alpha \Delta^\pm_\varepsilon (x) $. 
These figures, where the ridges show the highest transparency, illustrate the similarity of the tunneling behavior. At small $k$, the ridges are located exactly on the sets $\Sigma^+$ and $\Sigma'$ (Fig.~\ref{fig.3}), and $\Sigma^-$ and $\Sigma'$ (Fig.~\ref{fig.4}).
As further demonstrated by Figs.~\ref{fig.3} and \ref{fig.4}, for low particle energies, any dependence on $k$ is absent, but with increasing $k$, the ridges  shift to smaller $\alpha$ compared with the solutions to Eqs.~(\ref{17})-(\ref{21}). The central region corresponds to overbarrier transmission which is obviously transparent. The subbarrier tunneling is found in the region bounded above by the curves: 
\begin{equation}
k < \left\{ \begin{array}{ll} 
\sqrt{2\alpha \over c_1 (c_1 +c_2)} \, \varepsilon^{-1}
 & \mbox{for} ~~ \alpha >0 , \\
\sqrt{2 |\alpha | \over c_2 (c_1 +c_2)} \, \varepsilon^{-1}
& \mbox{for} ~~ \alpha < 0.
\end{array} \right.
\label{31}
\end{equation}

\begin{figure}
%\begin{center}
%\begin{center}
\centerline{\includegraphics[width=0.5\textwidth]{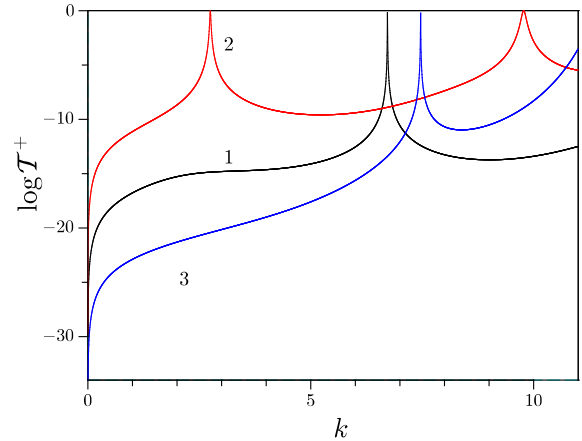}}
%\end{center}
\vspace{2pt}
\caption{ (Color online) Three sections of surface ${\cal T}^+(\alpha, k)$ at fixed strengths: $\alpha = 25$ (curve 1, black), $\alpha = -16$ (curve 2, red) and $\alpha = -34$ (curve 3, blue). The parameters chosen for calculations are the same as for Fig.~\ref{fig.3}. }
%\end{center}
\label{fig.5}
\end{figure}

In order to illustrate the profile of the ridges described by Figs.~\ref{fig.3} and \ref{fig.4}, we have plotted the dependence of the amplitude ${\cal T}^+$ on the wave number $k$ at some fixed values of $\alpha$ being close to the solutions of Eqs.~(\ref{17}) and (\ref{21}). We have restricted ourselves to the case with total transmission and found that the corresponding peaks are very narrow in the natural scale. However, in the logarithmic scale, the peaks essentially spread and the profiles become much more convenient for visualization as demonstrated by 
Fig.~\ref{fig.5}. 
\begin{figure}
%\begin{center}
%\begin{center}
\centerline{\includegraphics[width=0.5\textwidth]{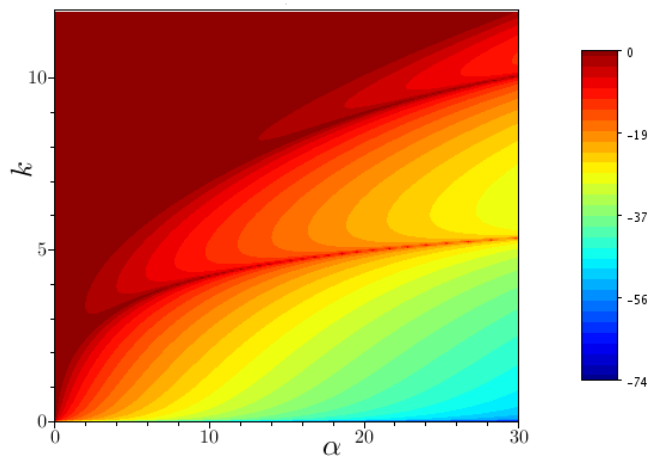}}
%\end{center}
\vspace{2pt}
\caption{(Color online) Transmission probability ${\cal T}^-$ plotted as a function of the strength $\alpha$ and the wave number $k$ with $\sigma =0$. The rest of parameters chosen for calculations are the same as for Fig.~\ref{fig.4}. }
%\end{center}
\label{fig.6}
\end{figure} 
The peaks in this figure correspond to the ridges depicted in Fig.~\ref{fig.3}. Note that these peaks correspond to the quantized values of  strengths: $\alpha= \alpha_{-3}^+,~\alpha_{-2}^+, ~\alpha_2^+$ at which the total transmission takes place.   

It is of interest to clarify the origin of the resonances described above for the potentials $\Delta_\varepsilon^\pm(x)$ with $\sigma =1$. To this end, we have compared these resonances with the well known phenomenon of the tunneling through DB systems. The DB structure can immediately be obtained from both the potentials $\alpha \Delta^\pm_\varepsilon(x)$ if we remove the wells, i.e., put $\sigma =0$ in Eqs.~(\ref{5})-(\ref{7}). For instance, in the case of the potential $\alpha \Delta^-_\varepsilon(x)$ where $\lambda_{11}^- = \lambda_{22}^-$, the requirement 
$v^- = k\lambda^-_{12} + k^{-1} \lambda^-_{21} =0$ results in the well-known resonance condition 
\begin{equation}
\left( {p \over k} + {k \over p} \right) \tan(pl) = 2 \cot(2kr) .
\label{32}
\end{equation}

On the other hand, as follows from the form of Eqs.~(\ref{17})-(\ref{21}), all the solutions to these equations run along the $\alpha$-axis to infinity and they obviously cannot be obtained from Eq.~(\ref{32}). Consequently, one can conclude that the presence of wells adjacent to the barriers makes these combined structures transparent at some quantized values of the strength $\alpha$. In this regard, we can refer this type of transmission to as a BW resonant tunneling. One can suppose that this tunneling differs in its nature from the double barrier resonant transmission. This difference can be better understood if we compare the contour plots given by Figs.~\ref{fig.4} ($\sigma =1$) and \ref{fig.6} ($\sigma =0$), where the rest of the parameters are the same.   
As shown in these figures, the ridges of resonant tunneling are ``perpendicularly'' arranged with respect to each other. Under the well presence, the resonances occur nearby the quantized values of the parameter $\alpha$ being very sensitive while varying this parameter. In the case of the well absence, the $\alpha$-dependence of the resonances disappears at all. 

\section{Conclusions and discussion}

Although rectangular systems are mainly straightforward exercises in textbooks on quantum mechanics, nevertheless, their various parametrizations leading to one-dimensional point or contact interactions provide rich objects with non-trivial properties. In this paper, we have introduced a new type of single point interactions with a fully resonant transparency. Their finite-range BW analogues are not necessary to be well-defined distributions in the squeezing limit. However, in this limit, the corresponding total Hamiltonian can be defined as a self-adjoint operator. The significant difference from the other point interactions studied so far is the appearance of the cancellation of divergences in the zero-range limit, namely, the cancellation of singularities emerging from the barrier and the well terms resulting in the existence of discrete ``allowed'' values of potential strengths.

More precisely, using the approach based on point interactions, a new type of tunneling through thin structures, which are composed from adjacent barriers and wells, has been observed. As particular examples, the two rectangular models with the potentials $\alpha \Delta^\pm_\varepsilon (x)$, which are schematically shown in Fig.~\ref{fig.1}, have been studied both analytically and numerically. For each of these models, a pair of equations with respect to the potential strength constant $\alpha$ and the ``well'' parameters $\sigma^\pm$ has been derived in the $\varepsilon \to 0$ limit: Eqs.~(\ref{17}) and (\ref{19}) for $\alpha \Delta^+_\varepsilon (x)$, and Eqs.~(\ref{19}) and (\ref{21}) for 
$\alpha \Delta^-_\varepsilon (x)$ permitting three discrete sets of solutions 
 $\Sigma^\pm $ and $\Sigma' $.
The potential $\alpha \Delta^+_\varepsilon (x)$ is fully transparent on the set $\Sigma^+$,
whereas on the set $\Sigma'$ it admits only a partial transmission as clearly demonstrated by Fig.~\ref{fig.2}. Contrary, due to the spatial symmetry, the potential $\alpha \Delta^-_\varepsilon (x)$ is fully transparent on both the sets 
$\Sigma^-$ and $\Sigma'$. Outside these three sets, both the potentials are opaque acting as a fully reflecting wall.  

As follows from Eqs.~(\ref{17})-(\ref{21}), the admissible values of the potential strength constant $\alpha$ do not depend on the energy of an incident particle. This seems to be unusual, however, as illustrated by Figs.~\ref{fig.3} and \ref{fig.4}, the resonant dependence on $k$ appears, when the potential parameters are finite. 
For better understanding the role of adjacent wells in the resonant tunneling, we have introduced the parameters $\sigma^\pm = \sigma,~0 < \sigma < \infty$, into the potentials $\alpha \Delta^\pm_\varepsilon (x)$, being responsible for the well depth, so that $\sigma^-$ ($\sigma^+$) controls the well potential if $\alpha >0$ 
($\alpha <0$) as defined below Eqs.~(\ref{3}). 
In the $\sigma \to 0$ limit, the potentials $\alpha \Delta^\pm_\varepsilon (x)$ are transformed to a typical DB structure. While performing this limit, the solutions to Eqs.~(\ref{17})-(\ref{21}) continuously run to infinity and Figs.~\ref{fig.4} and \ref{fig.6} clearly illustrate the transmission difference between the cases  $\sigma =1$ and $\sigma =0$. Vice versa, when one or two wells appear, i.e., $\sigma $ starts to increase from zero, the right tails of the ridges shown in Fig.~\ref{fig.6} are approaching the $\alpha$-axis at infinity and with further increasing $\sigma$, the countable series of these tails runs to the origin $\alpha =0$ as $\sigma \to \infty$, so that the $(\alpha, \, k)$-relief is becoming deformed and ``rotated'', taking finally the form shown in Fig.~\ref{fig.4}. This behavior explains why it is impossible to obtain a point interaction from a DB potential (without wells).   

Concerning possible physical implications of the resonance properties described above, one can notice that according to  
 Eqs.~(\ref{17})-(\ref{21}), the location of the resonance sets $\Sigma^\pm$ and $\Sigma'$ is very sensitive while varying the well-potential part, i.e., the parameter $\sigma^-=\sigma$ if $\alpha >0$  ($\sigma^+ =\sigma$ if $\alpha >0$). Therefore it is reasonable in practice to design the barriers as system parameters given by the strength $\alpha$, whereas the wells as adjustable external potentials controlled by the parameter $\sigma$. Inspired by the recent work of Turek and Cheon \cite{tc}, one can adopt the well part as an external attractive potential controlled by $\sigma$. In other words, the role of the barrier potential part will play $\alpha$, whereas the well potential will be controlled by the parameter $\alpha \sigma$. Obviously, for a given $\alpha$ belonging to the sets $ \Sigma^\pm$ or $\Sigma'$, one can adjust a certain value of $\sigma$ at which the barriers will be fully transparent for the flow of quantum particles. Any deviation from this value will block up the flow completely. In this way, one can design a quantum device like a transistor controlling the flow of electrons through barriers by an external potential. The situation is very similar to a spectral filter designed in Ref.~\cite{tc} where a sharp resonance termed the ``threshold resonance'' has been observed when the energy of incident particles equals the applied external potential. Roughly speaking, due to the ``rotation'' of the resonance picture illustrated by 
Figs.~\ref{fig.4} and \ref{fig.6}, the wave number $k$ is replaced by the strength $\alpha$ and this is a main difference between the $\alpha$-resonances described in our case and the threshold $k$-resonances admitting the spectral filtering transmission.  

\begin{figure}
%\begin{center}
%\begin{center}
\centerline{\includegraphics[width=0.5\textwidth]{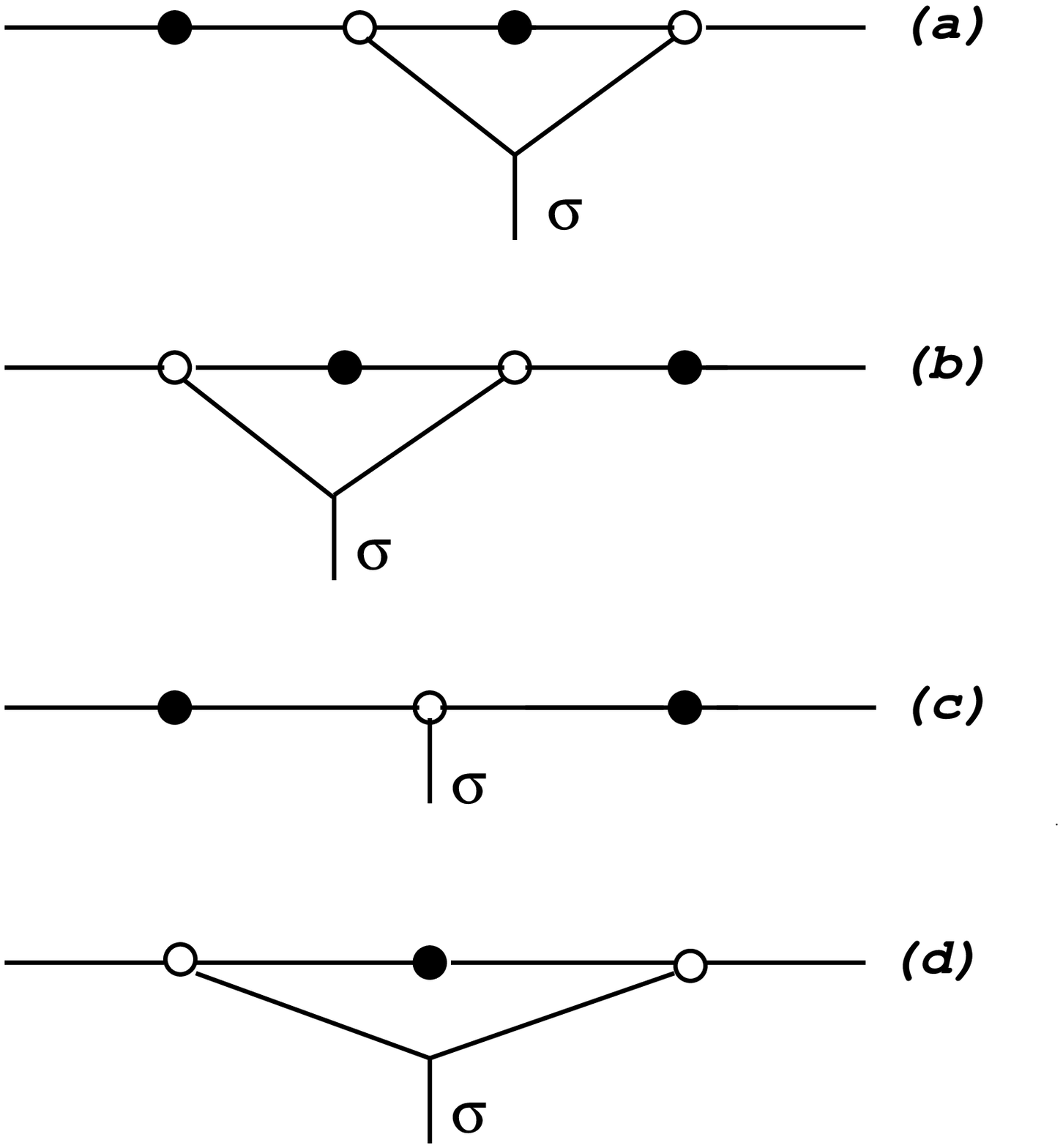}}
%\end{center}
\vspace{2pt}
\caption{ Graphs that correpond to the double $\delta'$-like point interactions obtained by squeezing the regular functions $\Delta_\varepsilon^\pm(x)$. Graphs (a) and (b) describe the point interaction obtained from the potential  $\alpha \Delta_\varepsilon^+(x)$ with $\alpha >0$ and $\alpha <0$, respectively. Similarly, graphs (c) and (d) correspond to the potential  $\alpha \Delta_\varepsilon^-(x)$ with $\alpha >0$ and $\alpha <0$, respectively. The filled balls mean barriers and the empty balls together with edges correspond to the external potentials controlling by parameter $\sigma$. 
}
%\end{center}
\label{fig.7}
\end{figure}

Next, similarly to the quantum Turek-Cheon device \cite{tc}, the BW structures in which the role of the well plays an external potential can schematically be depicted as quantum graphs. In our particular case, the graphs that correspond to Fig.~\ref{fig.1} are shown in 
Fig.~\ref{fig.7}. Here, the lines (a), (b), and (c) describe the device with two barriers shown by the filled balls which are connected through a terminal (shown by the empty ball) with an external potential controlled by the parameter $\sigma$. On 
the lines (a) and (b), the same potential is also applied at the side of one of the barriers. In fact, this is one device and the difference concerns its input and  output as well as the parameter interchanging $b \leftrightarrow b^{-1}$.  

\begin{figure}
%\begin{center}
%\begin{center}
\centerline{\includegraphics[width=0.5\textwidth]{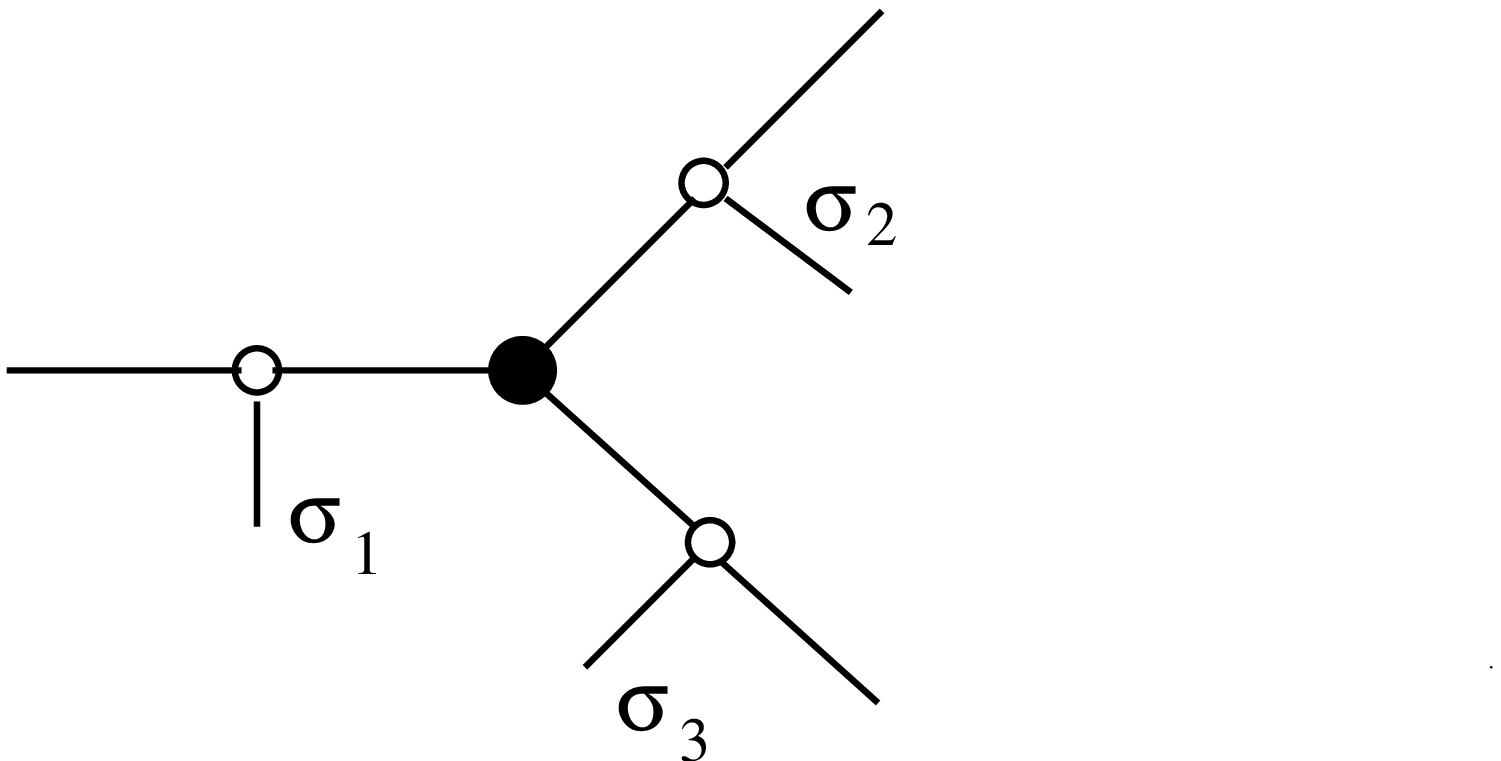}}
%\end{center}
\vspace{2pt}
\caption{ Graph with a vertex (filled ball), which describes a repulsive barrier junction, and three entering edges, at the endpoints of which terminals (empty balls) are placed where attractive potentials controlled by the parameters $\sigma_1$, $\sigma_2$, and $\sigma_3$ are switched on. 
}
%\end{center}
\label{fig.8}
\end{figure}

Of particular interest for resonant tunneling is a single barrier with two adjacent identical wells schematically shown in Fig.~\ref{fig.7}(d). First, the phenomenon of full resonant tunneling across a single barrier seems to be unusual, but the presence of two identical wells provides a full transmission at certain quantized values of the external parameter $\sigma$. Second, from a technological point of view such a device can be realized much easier than the others. A single barrier can be created simply by a natural contact (junction) of two quantum wires, whereas an external potential can be applied at the endpoints of edges entering this junction. By the term ``junction'' we mean a part of the system from which several wires emerge.
Varying $\sigma$, one can permit the quantum flow to pass across this junction totally or block it up completely. Obviously, this system with two terminals may be generalized to several terminals like that schematically illustrated by Fig.~\ref{fig.8}, where in general the parameters $\sigma_j, \, j=1,\, 2, \, 3$, are
different and any of them  may be assumed zero (the absence of wells).
 
Thus, 
 we have treated the two simplest examples of multilayer structures which are illustrated by Fig.~\ref{fig.1} and in the zero-range limit can be approximated by point interactions with total resonant tunneling. The tunneling effect observed in this paper by using the point interaction approach may be used in more realistic structures like a resonant tunneling diode with an additional well termed  ``prewell'' \cite{bbkl}. Graph (b) in Fig.~\ref{fig.7} corresponds to this device if the external potential is applied only to the edge in front of the DB structure.
Nevertheless, the structures with the well located between the barriers should be of interest as well because in this case one can get a full resonant tunneling. 
In general, the external potential parameters $\sigma_j, \, j=1, \, 2, \ldots, $ can be assumed different and similar methods can be used for studying more complicated geometries. We hope to pursue this point in the future.

The phenomenon described in this paper should be observable not only in quantum systems, but also in systems in which classical waves can propagate.
In view of the formal analogy between the Schr\"{o}dinger equation and the electromagnetic Helmholtz equation, this phenomenon can be directly applied to a medium with one-dimensional stepwise inhomogeneity of dielectric permeability. While the experiments involving electrons are usually difficult to realize (mainly due to the smallness of the electron de Broglie wavelength at usual temperatures), the observations on microwave tunneling provide clear data. Besides this, for instance, for TE waves  ($E_x =H_y=E_z =0$, an incident wave and inhomogeneity coincide with the $x$ direction), the analogue of the strength $\alpha$ appears to be the frequency of electromagnetic field which can be tuned \cite{ma}.

\begin{center}
{\bf  ACKNOWLEDGMENTS}
\end{center} 

The author acknowledges the financial support from the Ukrainian State Grant for Fundamental Research No.~0112U000056. He would like to express gratitude to 
Y.~Zolotaryuk for valuable suggestions and the help in numerical computations.

\end{document}